\documentclass[aps,prd,preprint,superscriptaddress,showkeys,nofootinbib]{revtex4-2}

\usepackage{natbib}

\usepackage{bookmark}
\usepackage{changes}
\usepackage{graphicx}
\usepackage{amsmath}

\usepackage{needspace}
\usepackage{setspace}
\usepackage{tabularx}

\usepackage{subcaption}
\usepackage{multirow}
\usepackage{MnSymbol}
\usepackage{xspace}

\usepackage{booktabs}
\usepackage{bigints}

\usepackage{tikz}
\usetikzlibrary{decorations.markings,calc}
\usetikzlibrary{patterns.meta}
\usepackage{circuitikz}
\usepackage{xcolor}

\usepackage{enumitem}
\setlist[itemize]{noitemsep}
\usepackage[parfill, indent]{parskip}
\usepackage{graphicx}

\usepackage[separate-uncertainty = true]{siunitx}

\usepackage{epstopdf}

\AtBeginDocument{\RenewCommandCopy\qty\SI}

\usepackage{hyperref}

\newcommand{\diff}[1]{\operatorname{d}\ifthenelse{\equal{#1}{}}{\,}{\!#1}}

\newcommand{\ecm}{\ensuremath{\si{\elementarycharge}\!\cdot\!\cm}}

\newcommand{\cm}{\ensuremath{\mathrm{cm}}}

\newcommand{\CP}{\ensuremath{C\hspace{-1pt}P}\xspace}

\begin{document}

\title{Experimental search for electric dipole moments of light
	radioactive nuclei}
\author{Chavdar~Dutsov\footnote{Present address: CERN, Esplanade des
		Particules 1, Meyrin, Switzerland}}
\email{chavdar.dutsov@cern.ch} \affiliation{PSI Center for Neutron and
	Muon Sciences, 5232 Villigen PSI, Switzerland}

\author{Timothy~Hume}
\affiliation{PSI Center for Neutron and Muon Sciences, 5232 Villigen
	PSI, Switzerland} \affiliation{ETH Zürich, 8093 Zürich, Switzerland}

\author{Maxim Pospelov}
\affiliation{School of Physics and Astronomy, University of Minnesota,
	Minneapolis, MN 55455, USA} \affiliation{ William I. Fine
	Theoretical Physics Institute, University of Minnesota, Minneapolis,
	MN 55455, USA}

\author{Philipp~Schmidt-Wellenburg}
\affiliation{PSI Center for Neutron and Muon Sciences, 5232 Villigen
	PSI, Switzerland}

\begin{abstract}
	We discuss a search for the electric dipole moment (EDM) of a light
	beta-radioactive ion using a compact ion trap by adapting the
	``frozen-spin'' method. The measurement will be done on ions
	stripped of their valence electrons, thereby bypassing the
	significant Schiff screening that hinders the application of
	successful contemporary EDM searches using heavy neutral atoms and
	molecules to light nuclei. %
	%
	%
	We identified $^8$Li as the most promising candidate for a
	proof-of-concept EDM search and we estimate that the current
	indirect proton EDM limit of a few $\SI{e-25}{\ecm}$ set by
	$^{199}$Hg measurements can be surpassed with a week of measurement
	time at existing facilities.
\end{abstract}

\maketitle 

\clearpage

\section{Introduction}

The search for electric dipole moments~(EDMs) represents one of the
most promising avenues in particle physics~\cite{Chupp2019} to reveal
the nature of beyond Standard Model~(SM) physics. A permanent EDM in
any fundamental particle or system would constitute direct evidence of
time-reversal ($T$) violation and, through the $C\hspace{-1pt}PT$
theorem~\cite{Luders1954,Tureanu2013}, would also violate
charge-parity (\CP) symmetry. While the Standard Model incorporates
\CP violation via the complex phase of the Cabibbo-Kobayashi-Maskawa
(CKM) matrix~\cite{Kobayashi:1973fv}, its magnitude is insufficient to
explain the observed matter-antimatter asymmetry through standard
electroweak baryogenesis~\cite{Hocker:2001jb,Pospelov:2013sca}. EDMs
thus provide a model-independent probe of new \CP violating physics
beyond the SM.\
Since the pioneering neutron EDM experiment by Purcell, Ramsey, and
Smith~\cite{Smith1957}, EDM searches have expanded to diverse systems:
neutrons, muons, paramagnetic atoms/molecules (Cs, Tl, YbF, ThO,
HfF$^+$), and diamagnetic atoms ($^{129}$Xe, $^{199}$Hg, $^{225}$Ra,
TlF)~\cite{Chupp2019}. Despite decades of measurements, all results
remain consistent with zero, placing stringent constraints on
\CP-violating new physics.

To date, no direct measurement of the proton EDM has been
accomplished. The leading experimental proposal involves an
all-electric storage ring designed to reach a sensitivity three orders
of magnitude beyond that of current neutron EDM searches. This
approach requires a storage ring with a circumference of approximately
\SI{800}{m}~\cite{pEDMarXiv, Rathmann2014, CPEDM_2021, Rathmann2020}.
A search for oscillating EDMs induced by axion-like particles,
demonstrated using magnetic storage of polarized deuterons at the
Cooler Synchrotron COSY (Research Center J\"ulich, Germany), is an
example of initial progress in this direction~\cite{Karanth2023}. 

So far hadronic \CP-violating interactions, particularly the proton
EDM, were probed through the nuclear Schiff moment of atoms that have
closed electron shells (diamagnetic)~\cite{Chupp2019}. The current
best limit on the EDM of such a system is provided by $^{199}$Hg:
\SI{7.4e-30}{\ecm} (95\% C.L.)~\cite{Graner2016}.

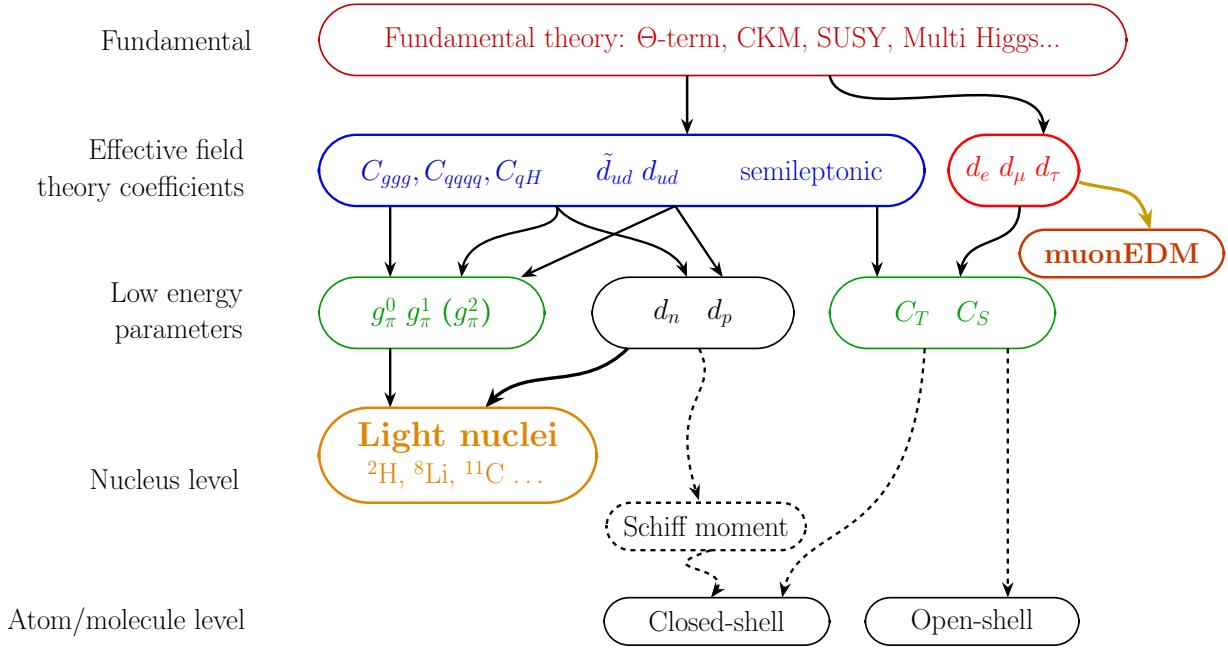
\begin{figure}[tphb]
	\centering \resizebox{1\textwidth}{!}{%
		\begin{circuitikz}
	\tikzstyle{every node}=[font=\Large]

	\draw [ color={rgb,255:red,194; green,0; blue,0} , line width=1pt ,
		rounded corners = 22.5] (9.75,22.5) rectangle node {Fundamental
			theory: $\Theta$-term, CKM, SUSY, Multi Higgs...} (26.75,21);
	\draw [ color={rgb,255:red,0; green,153; blue,3} , line width=1pt ,
		rounded corners = 22.5] (9.75,16.75) rectangle node
		{$g_\pi^0~g_\pi^1~(g_\pi^2)$} (14.5,15.25);
	\draw [ line width=1pt , rounded corners = 22.5] (15.5,16.75)
	rectangle node {$d_n\quad d_p$} (19.75,15.25);
	\draw [ color={rgb,255:red,0; green,158; blue,11} , line width=1pt ,
		rounded corners = 22.5] (20.5,16.75) rectangle node { $C_T\quad
				C_S$} (25.25,15.25);
	\draw [ line width=1pt , rounded corners = 15.0] (15.75,10) rectangle
	node { Closed-shell} (20.5,9);
	\draw [ line width=1pt , rounded corners = 15.0] (21.25,10) rectangle
	node { Open-shell} (25.75,9);
	\draw [line width=1.4pt, ->, >=Stealth] (11.25,18.25) --
	(11.25,16.75);
	\draw [ color={rgb,255:red,0; green,7; blue,214} , line width=1.4pt ,
		rounded corners = 22.5] (9.75,19.75) rectangle node { $C_{ggg}, C_{qqqq}, C_{qH} \qquad \tilde d_{ud}~d_{ud}\quad\quad$
	semileptonic} (22.5,18.25);
	\draw [line width=1.4pt, ->, >=Stealth] (14.75,18.25) .. controls
	(15,17.5) and (12.75,18) .. (12.75,16.75) ;
	\draw [line width=1.4pt, ->, >=Stealth] (14.75,18.25) .. controls
	(15.25,17.5) and (17.25,17.75) .. (17.5,16.75) ;
	\draw [line width=1.4pt, ->, >=Stealth] (17.25,18.25) -- (14,16.75);
	\draw [line width=1.4pt, ->, >=Stealth] (17.25,18.25) --
	(18.25,16.75);
	\draw [line width=1.4pt, ->, >=Stealth] (21.5,18.25) -- (21.5,16.75);
	\draw [line width=1.4pt, ->, >=Stealth] (11.25,15.25) -- (11.25,14);
	\draw [line width=2pt, ->, >=Stealth] (16.25,15.25) .. controls
	(15.25,14.25) and (14,15) .. (13.25,14) ;
	\draw [line width=1.4pt, ->, >=Stealth, dashed] (18,11) .. controls
	(16.5,10.5) and (18.75,10.75) .. (18,10);
	\draw [line width=1.4pt, ->, >=Stealth, dashed] (24.25,15.25) --
	(24.25,10.00);
	\draw [line width=1.4pt, ->, >=Stealth, dashed] (22.5,15.25) ..
	controls (22.25,10.75) and (19.75,11.75) .. (19.5,10);
	\draw [line width=1.4pt, ->, >=Stealth, dashed] (17.75,15.25) ..
	controls (18.25,14.25) and (17,14.25) .. (17.75,12.05);
	\draw [ line width=1.4pt , rounded corners = 14.8, dashed] (15.8,12)
	rectangle node {Schiff moment} (20,11);

	\draw [line width=1.5pt, ->, >=Stealth] (17.5,21) -- (17.5,19.75);
	\draw [ color={rgb,255:red,255; green,0; blue,0} , line width=1.5pt ,
		rounded corners = 22.5] (23,19.75) rectangle node {
			$d_e~d_\mu~d_\tau$} (25.75,18.25);
	\draw [line width=1.5pt, ->, >=Stealth] (20.5,21) .. controls
	(20.75,19.75) and (25.25,21.5) .. (25,19.75) ;
	\draw [line width=1.5pt, ->, >=Stealth] (24.5,18.25) .. controls
	(24.5,17.25) and (23.25,17.75) .. (23.25,16.75) ;
	\draw [ color={rgb,255:red,224; green,131; blue,0} , line width=1.5pt
		, rounded corners = 30.0] (9.75,14) rectangle node[align=center]
		{\LARGE \textbf{Light nuclei}\\ $^2$H, $^8$Li, $^{11}$C \ldots}
	(15.6,12);
	\draw [ color={rgb,255:red,199; green,56; blue,0} , line width=1.5pt ,
		rounded corners = 15.0] (24.5,17.75) rectangle node
		{\textbf{muonEDM}} (28.8,16.75);
	\draw [ color={rgb,255:red,199; green,156; blue,0}, line width=2pt,
		->, >=Stealth] (25.75,18.75) .. controls (26.75,18.5) and (27,18.75)
	.. (27.25,17.75) ;
	\node [font=\Large] at (6.5,12.5) {Nucleus level};
	\node [font=\Large] at (5.7,9.5) {Atom/molecule level};
	\node [font=\Large] at (6.75,21.75) {Fundamental};
	\node [font=\Large, align=right] at (6.75,16) {Low energy\\
		parameters};
	\node [font=\Large, align=right] at (6,19) {Effective field\\ theory
		coefficients};
\end{circuitikz} }%

	\caption{Illustration of the connection between underlying
		fundamental theories to laboratory measurements, passing through
		potential effective \CP-violating sources. Note that measurements
		on fully or partially stripped ions have a more direct and
		unambiguous connection to low-energy parameters than
		atoms/molecules. The muonEDM experiment~\cite{Adelmann2025}, which
		serves as a blueprint for the light ion search, would measure the
		EDM of an elementary particle providing a direct connection to
		fundamental theories. Figure adapted from~\cite{Chupp2019}.}
	\label{fig:theory_experiment_relations}
\end{figure}

Despite the exceptional experimental sensitivity, the implications of
this result are somewhat reduced due to the large Schiff screening of
the nuclear EDM in neutral atoms~\cite{Schiff1963}. Furthermore,
extracting fundamental hadronic EDMs from atomic measurements involves
a elaborate chain of nuclear and atomic many-body theory, as depicted
in Fig.~\ref{fig:theory_experiment_relations}.

The $^{199}$Hg EDM measurement is interpreted in terms of a Schiff
moment, which -- under simplified nuclear structure assumptions and
theoretical treatment~\cite{Dmitriev2003} -- translates into a
constraint on the proton EDM of approximately
\SI{2.0e-25}{\ecm}~\cite{Graner2016}, if it were the sole contributor
to the overall EDM.

We propose an alternative search for hadronic \CP violation using
beta-decaying, fully stripped or closed-shell light ions confined in a
compact ion trap. The partial or complete removal of electrons allows
the nucleus to interact directly with an external electric field,
substantially mitigating atomic screening and enabling a more direct
measurement of the effective nuclear EDM\@.

As we will demonstrate, sensitivities competitive with or superior to
current hadronic EDM limits can be achieved using a compact setup.
Unlike the large-scale infrastructure required for storage-ring
experiments, the small anomalous magnetic moments~(AMM) of selected
light ions enable EDM measurements to be performed in a
centimeter-scale electromagnetic trap. Our proposed configuration
employs an $\sim\SI{40}{keV}$ ion beam confined in a closed magnetic
orbit, producing otherwise more difficult to achieve electric fields
of the order of \SIrange{1}{10}{MV/m} in the ion rest-frame.

\begin{figure}[h]
	\centering \includegraphics[]{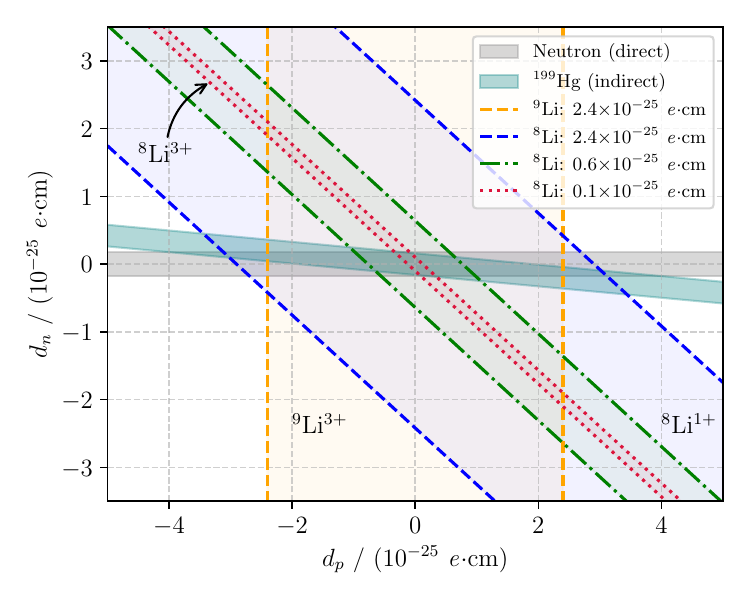}
	\caption{ Constraints in the $d_n$--$d_p$ parameter space. Light
	blue and gray are the existing constraints on the mercury and
	neutron EDMs. The mercury EDM band includes a $\pm 2\sigma$
	theoretical uncertainty. The intersection of these two bands sets
	the limit on the proton EDM, $|d_p|<4\times 10^{-25}\,e{\rm cm}$.
	Projected sensitivities from $^8$Li EDM measurements are shown
	assuming: one week of data collection (blue dashed line),
	\SI{100}{days} of data (green dash-dotted line), and \SI{7}{days}
	with fully stripped $^8$Li$^{3+}$ ions (red dotted line). Even
	with a one-week measurement, the $^8$Li EDM constraint improves
	existing bounds on $d_p$.}
	\label{fig:li8_pedm_pdf}
\end{figure}

The isotope $^8$Li in particular serves as an excellent probe due to
the very low AMM of the fully stripped state, the high abundance in
isotope production facilities, and the fact that it is a standard
beta-NMR probe~\cite{Gins2019}. Moreover, it has unpaired neutron and
proton spins, which leads to a balanced sensitivity to both neutron
and proton EDMs ($d_n$, $d_p$) in contrast to the $^{199}$Hg~atom,
which is predominantly sensitive to the neutron EDM, as shown in
Fig.~\ref{fig:li8_pedm_pdf}. If it can be measured to the same
sensitivity, the fully stripped $^9$Li is also interesting as a
predominantly proton EDM probe.

With an estimated experimental sensitivity of $\SI{2.3e-25}{\ecm}$, a
one-week measurement of $^8$Li or $^9$Li EDM would significantly
improve constraints on $d_p$, especially in a scenario where neutron
and mercury EDM bounds are nearly parallel in the $d_n$–$d_p$ plane.
This makes lithium isotopes particularly attractive for testing new
\CP violation mechanisms and serves as a stepping stone to more
complex nuclear systems with potentially enhanced \CP-violating
effects.


The direct search for an EDM of a bare nucleus or a partially stripped
ion probes nuclear and hadronic sources of \CP violation free of large
screening effects. Focusing on $^8$Li we will pioneer methods to probe
EDMs of light nuclei allowing different sensitivity to specific
fundamental sources of \CP violation -- a necessary addition
considering that a global analysis of multiple searches is needed to
unambiguously constrain new physics parameters~\cite{Chupp2015}.

\section{Methodology}

Farley, Semertzidis, Khriplovich and
others~\cite{Khriplovich1998,Semertzidis2001,Farley2004PRL} proposed a
method to measure EDMs of charged systems -- muons in particular -- in
storage rings, known as the frozen-spin technique. The essence of the
frozen-spin technique is the cancellation of the anomalous $(g-2)$
precession by applying a radial electric field perpendicular to the
momentum of the stored particles and to the magnetic field. That way
any remaining precession is a consequence of the EDM, which manifests
itself through a precession of the spin around the electric-field
vector in the particle's rest frame.

One major advantage of using a storage ring is that the motional
electric field experienced by the particle can exceed by far any
static field achievable in a laboratory. Furthermore, the frozen-spin
technique enables a linear build-up of the EDM signal over time, which
is otherwise limited by the much shorter anomalous spin precession
period. The frozen-spin muonEDM experiment at the Paul Scherrer
Institute (PSI, Switzerland) represents the first attempt to implement
this technique, aiming to achieve unprecedented sensitivity in muon
EDM measurements~\cite{SchmidtWellenburg2023c, Sakurai2022,
	Adelmann2021arXiv, SchmidtWellenburg2023b, Adelmann2025}.

The electric field strength required to freeze the spin is
proportional to the AMM $a = (g-2)/2$, the beam velocity $\vec v =
	\vec \beta c$ and the magnetic field $B$, while the EDM sensitivity is
proportional to $\lvert \vec v \times \vec B \rvert$. Therefore, the
maximum achievable sensitivity is ultimately constrained by the
practical limits of the electric field needed to freeze the spin.
Systems with smaller AMM are thus advantageous. For example, the muon,
as an elementary particle, has a relatively small AMM of approximately
\num{1.17e-3}.

Although ions do not generally exhibit such small AMM, certain ions
possess an effective $g$-factor near 2, making it possible to apply
the frozen-spin technique with more modest electric fields.
Khriplovich~\cite{Khriplovich1998, Khriplovich2002} demonstrated this
for isotopes with mass number $A > 24$, where ions with AMM similar to
the muon can be found, however all highly charged.

Here we extend this approach to lighter nuclei with $A<24$,
particularly $^8$Li, which offers several advantages for a frozen-spin
experiment. As a well established isotope for beta-NMR, $^8$Li is
abundantly produced at isotope production facilities such as TRIUMF
(Canada) and ISOLDE (CERN). Furthermore, its relatively simple nuclear
structure allows for precise theoretical treatment from first
principles.

\subsection{Measurement principle}

\begin{figure}[htb]
	\centering \includegraphics[width=\textwidth]{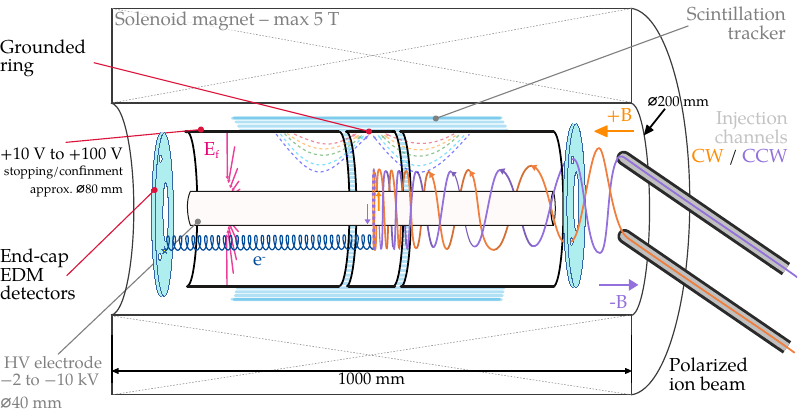}
	\caption{\protect\rule{0ex}{18pt}Schematic cross-section of the
		experimental apparatus (components not to scale). Grayed-out
		components are already existing in the muonEDM experiment and
		those indicated in red lines are the new developments needed for
		storing light ions: an electrode system for stopping and storing
		ion bunches propagating clockwise~(CW) or counter-clockwise~(CCW)
		after injection through magnetically shielded channels, and
		scintillation beta-decay detectors. }
	\label{fig:psc_setup}
\end{figure}

Figure~\ref{fig:psc_setup} shows a schematic of the experimental setup
of the proposed light ion EDM (LionEDM) spectrometer. We propose to
measure the EDM of beta-radioactive ions using a modified version of
the compact frozen-spin trap developed for the measurement of the muon
EDM at PSI~\cite{Adelmann2025}. The apparatus will be located at the
exit of the beamline, where spin-polarized ions with kinetic energy
$\simeq$\SI{42}{keV} will travel through a magnetically shielded
injection channel into the solenoid.

Achieving high sensitivity requires an intense source of polarized ion
beams, such as those available at the VITO beamline at ISOLDE,
CERN~\cite{Gins2019}. At this facility, polarization is transferred
from the electron shell to the nucleus via finely tuned laser
excitation of the hyperfine structure~\cite{Bergmann2002}.

As the ions enter the magnetic field, they will spiral inward with an
orbit radius of approximately $R \approx \SI{30}{mm}$. A timed
electric pulse will decelerate them axially, allowing trapping within
an electrostatic potential well. The timing of this pulse will be
synchronized with the accelerator signal, eliminating the need for an
entrance trigger. The pulse, generated by applying an electric field
gradient between electrodes separated by \SI{2}{mm}, on each side of
the ground ring in the center, creates an axial electric field. Monte
Carlo particle simulations indicate a pulse duration in the range of
\SIrange{1}{3}{\micro\second}.

Once trapped, the ions will undergo both cyclotron motion and Larmor
precession due to the interaction between their spin and the magnetic
field. The difference between these two motions is also known as the
anomalous or $(g-2)$ spin precession. To precisely control this
effect, a radial electric field is applied within the storage volume.
Its strength is carefully tuned to cancel the anomalous spin
precession, effectively reducing the total spin precession frequency
to zero in the absence of an EDM\@.

Like muons, beta-decaying nuclei are ``self-analyzing,'' meaning their
average spin orientation can be measured by tracking the decay
products. Parity violation in weak decay leads to a preference for
high-energy electrons to be emitted in the direction of the spin. The
EDM will be determined from the change in asymmetry,
$d\mathcal{A}/dt$, where $\mathcal{A} = (N_\uparrow -
	N_\downarrow)/(N_\uparrow + N_\downarrow)$ represents the difference
between the number of electrons emitted parallel or antiparallel to
the main magnetic field.

The beta-decay asymmetry $\mathcal{A}$ will be measured using end-cap
detectors, positioned upstream and downstream along the magnetic field
direction, and a scintillation tracker positioned around the ion
orbit. The $(g-2)$ precession frequency will be deduced from the
oscillating number of detected decay products in the outer
scintillation tracker, similar to the beta-NMR technique.

Once the ion spin is ``frozen,'' the ion EDM $d$ will be determined
from the change of asymmetry in the end-cap detectors
\begin{equation}
	\frac{d\mathcal{A}}{dt} = \frac{2}{\hbar}\alpha P d E^*
	\label{eq:edm_calculation}
\end{equation}
where $E^* \approxeq c\beta B + E$ is the electric field strength in
the rest frame of the particle, $\alpha$ is the directional asymmetry
in the beta-decay, typically ranging from 0.33 to~1 depending on the
spin-parity configuration of the initial and final nuclear states, and
$P$ is the mean polarization of the ion beam.

\subsection{The frozen-spin technique}

The frozen-spin technique is central to the experiment, and we will
examine it in more detail here. The relativistic spin motion of a
particle moving through electric field $\vec E$ and magnetic field
$\vec B$ is described by the extended Thomas-BMT
equation~\cite{Thomas1926,Thomas1927,PhysRevLett.2.435},
\begin{equation}
	\vec{\omega} = \vec{\omega_a} + \vec\omega_e =
	-\frac{q}{m}\left[a\vec{B}-\left(a+\frac{1}{1-\gamma^2}\right)
		\frac{\vec{\beta}\times\vec{E}}{c}\right] -\frac{\eta
		q}{2m}\left[\vec{\beta}\times\vec{B}+\frac{\vec{E}}{c}\right]\,,
	\label{eq:omegaMuWithEDMsimple}
\end{equation}
considering a storage orbit where the particle momentum, magnetic
field, and electric field are mutually perpendicular, and adding a
term describing the effect of the EDM\@. Here
$\vec{\beta}=\vec{p}c/\mathcal{E}$ and
$\gamma=\left(1-\beta^2\right)^{-1/2}$ are the relativistic factors
for a particle with total energy $\mathcal{E}$, $a$ is the AMM, and
$\eta$ is the dimensionless constant which relates the spin to the
EDM\@, $\vec{d}=\eta q/(2mc)\vec{s}$.

The anomalous precession frequency $\vec \omega_a$ is the difference
between the Larmor precession and the cyclotron precession. The second
term $\vec\omega_e$ arises from the EDM coupling to the electric field
in the boosted reference frame of the moving particle. For particles
with spin $J > 1/2$, Eq.~\eqref{eq:omegaMuWithEDMsimple} contains
higher order terms with decreasing importance.

By adjusting the electric and magnetic field such that
\begin{equation}
	a\vec{B} = \left(a+\frac{1}{1-\gamma^2}\right)\frac{\vec{\beta}\times\vec{E}}{c},
	\label{eq:FrozenSpinCondition}
\end{equation}
it is possible to nullify the $(g-2)$ precession, $\omega_a=0$. The
frozen-spin technique achieves this cancellation of the anomalous
precession term by applying an electric field of magnitude $E_{\rm f}
	\approx a c \beta \gamma^2 \vert \vec B \vert$. In the absence of an
EDM, the spin orientation relative to the particle momentum remains
unchanged, ``frozen,'' during storage.

One approach to fulfill condition~\eqref{eq:FrozenSpinCondition} and
freeze the spin is to use an all-electric storage ring ($B = 0$) and
tune the particle momentum such that $\gamma^2 = (a+1)/a$. This
``magic'' momentum is then $p = mc/\sqrt{a}$, where $m$ is the
particle mass. It results in GeV-scale momentum requiring a storage
ring with hundreds of meters circumference for electrostatic storage.
This approach is only feasible for particles with relatively large $a$
and relatively low mass, and is considered for searches for the proton
and deuteron EDMs~\cite{Anastassopoulos2016, Karanth2023}.

In our approach, we opt for a combined electric and magnetic
confinement which allows for a very compact storage orbit. However, a
large value of $a$ would require excessively strong electric fields,
making the frozen-spin technique impractical. Fortunately, for ions,
this limitation can be mitigated by adjusting the ion charge $z$,
which fine-tunes the ion’s AMM~\cite{Khriplovich1998,
	Khriplovich2002}. This flexibility is extremely advantageous, as it
enables the use of ions with very small magnetic anomalies,
significantly enhancing the sensitivity of the EDM measurement for a
given isotope.

\subsection{Sensitivity to the ion EDM}

To be sensitive to the nuclear spin, we select ions that are fully
stripped or have only paired electrons in their shell. The favorable
numbers of electrons are $Z-z = [0, 2, 4, 6, 10, 12, 14, 18, \ldots]$
corresponding to an integer number of electron pairs in the electronic
shell. Ions with 8 and 16 electrons are omitted as they correspond to
$p^4$ orbitals for which two of the electrons are not paired.


Starting from the equation governing the time evolution of angular
momentum in the presence of a magnetic field,
\begin{equation}
	\frac{d\vec J}{dt} = \vec \mu \times \vec B = g \frac{q}{2M}\vec J
	\times \vec B,
	\label{eq:magnetic_moment}
\end{equation}
where $\vec J$ is the angular momentum of an ion with mass $M$ and
charge $q = ze$, we express the $g$-factor of an ion whose electron
shell contains only paired electrons as
\begin{equation}
	g = \frac{\lvert \vec \mu \rvert}{\lvert \vec J \rvert}
	\frac{2M}{q}.
	\label{eq:g_factor}
\end{equation}
Here, the nuclear magnetic moment is given by $\mu = \mu_m\mu_N$,
where $\mu_m$ is the measured magnetic moment of the nucleus in units
of nuclear magnetons, with $\mu_N = \frac{e\hbar}{2m_p}$. Substituting
this definition into the expression for $g$, we obtain
\begin{equation}
	g = \frac{\mu_m}{J} \frac{e}{2m_p} \frac{2M}{q} =
	\frac{\mu_m M}{zm_p J} \approxeq \frac{\mu_m}{J}\frac{A}{z},
	\label{eq:g_factor_deriv}
\end{equation}
where $m_p$ is the proton mass and $A$ is the atomic number of the
isotope. An exact formula takes into account the mass excess $\Delta$
and the number of missing electrons in the atomic shell $zm_e$ for the
ion mass,
\begin{equation}
	M = Am_u + \Delta - zm_e,
	\label{eq:ion_mass}
\end{equation}
where $m_u$ is the atomic mass constant, $m_u/m_p \approxeq
	0.992776098$ and $m_e$ is the electron mass. Combining
Eqs.~\eqref{eq:g_factor_deriv} and~\eqref{eq:ion_mass}, the exact
$g$-factor is
\begin{equation}
	g = 
	\frac{\mu_m}{J}\frac{A}{z}\left(\frac{m_u}{m_p} +
	\frac{\Delta}{Am_p} - \frac{zm_e}{Am_p} \right).
	\label{eq:exact_g_factor}
\end{equation}
The AMM $a = g/2 - 1$ is then
\begin{equation}
	a = \frac{\mu_m}{2m_pJ}\frac{A m_u + \Delta - zm_e }{z} - 1.
	\label{eq:magnetic_anomaly}
\end{equation}
For an ion with atomic number $Z$ and $Z-z$ paired electrons the
sensitivity to the EDM is
\begin{equation}
	\sigma(d_i) = \frac{\hbar}{2\tau\alpha P E_\mathrm{f}}
	\frac{a}{(a+1)} \frac{Z}{z} \frac{1}{\sqrt{N}},
	\label{eq:stat_sens_per_p}
\end{equation}
under the non-relativistic assumption, valid for the energy scale of
the experiment. The factor of $Z/z$ comes from the shielding of the
nuclear EDM by the electrons in the shell. The mean free spin
precession time due to an EDM is $\tau$ which is limited by the
lifetime of the isotope or, for longer lived nuclides, by the spin
coherence time. 

Taking into account that not all nuclei will decay within the
measurement time, the number of detected decay products is
\begin{equation}
	N = Y \varepsilon \left(1 - 2^{-\frac{\tau}{T_{1/2}}}\right)
	\frac{D}{\tau},
	\label{eq:number_decayed}
\end{equation}
where $Y$ is the yield in ions per bunch, $\varepsilon$ is the
trapping efficiency, $T_{1/2}$ is the isotope's half-life and $D$ is
the total duration of the experiment, such that $D/\tau$ represents
the number of injected ion bunches. The factor in brackets incurs a
penalty for long-lived nuclei where only a fraction of the injected
ions would decay during the measurement time. 

From Eq.~\eqref{eq:stat_sens_per_p} we can see that to maximize the
EDM sensitivity, for a given maximum achievable frozen-spin electric
field $E_{\mathrm{f}}$, we require an ion with low AMM $a$, available
in a high ionization state $z$ and in large quantities $N$.

To identify the optimal ions for EDM searches, we compiled recommended
datasets on isotope properties from the Evaluated Nuclear Structure
Data File (ENSDF)~\cite{ENSDF} including mass, nuclear spin, and
half-life, and supplemented them with nuclear magnetic moment
measurements~\cite{Mertzimekis2016, Mertzimekis2016_technical_report}.
We then filtered for beta-decaying ions that are either fully stripped
or have an integer number of electron pairs and calculated their AMM.

\setlength{\tabcolsep}{6pt} \begin{table}
	\caption{Isotope properties summary ordered by increasing anomalous
		magnetic moment. The branching ratio refers to the highest
		probability transition corresponding to the given spin-parity of the
		final state $J^{\pi'}$. All nuclides undergo beta-decay with 100\%
		probability. The mean beta-decay energy $\tilde E_\beta$ is also
		shown. The production yield per \textmu C proton beam on target is
		taken from the ISOLDE database~\cite{isoldeyields, Ballof2020}.}
	\label{tab:isotope_summary}
	\begin{tabular}{@{}lcccccccccc@{}}
		\toprule
		Ion                                 & $\mu$, $\mu_N$            & $a$                       &                               & $J^\pi \rightarrow J^{\pi'}$                  & Branching                   & $T_{1/2}$, s              & $\tilde E_\beta$, keV    & Yield, \textmu C$^{-1}$               \\
		\midrule
		$^{13}$O$^{6+}$                     & 1.389                     & -0.002                    & $\beta^+$                     & $3/2^-$ $\rightarrow$ $1/2^-$                 & 89.2\%                      & 0.009                     & 8136                     & ---                                   \\
		$^{18}$N$^{3+}$                     & 0.327                     & -0.024                    & $\beta^-$                     & $1^-$ $\rightarrow$ $1^-$                     & 47.2\%                      & 0.619                     & 4485                     & $6.5 \cdot 10^{1}$                    \\
		$^{19}$O$^{6+}$                     & 1.532                     & -0.036                    & $\beta^-$                     & $5/2^+$ $\rightarrow$ $3/2^+$                 & 54.4\%                      & 26.880                    & 1442                     & $1.3 \cdot 10^{5}$                    \\
		$^{9}$C$^{4+}$                      & 1.391                     & 0.039                     & $\beta^+$                     & $3/2^-$ $\rightarrow$ $3/2^-$                 & 54.1\%                      & 0.127                     & 7502                     & $4.0 \cdot 10^{3}$                    \\
		$^{22}$F$^{7+}$                     & 2.694                     & 0.051                     & $\beta^-$                     & $4^+$ $\rightarrow$ $4^+$                     & 53.9\%                      & 4.230                     & 2449                     & $2.5 \cdot 10^{4}$                    \\
		$^{17}$C$^{4+}$                     & 0.758                     & 0.067                     & $\beta^-$                     & $3/2^+$ $\rightarrow$ $1/2^+$                 & 27.0\%                      & 0.191                     & 5413                     & $8.0 \cdot 10^{0}$                    \\
		$^{12}$N$^{3+}$                     & 0.457                     & -0.091                    & $\beta^+$                     & $1^+$ $\rightarrow$ $0^-$                     & 96.2\%                      & 0.011                     & 7923                     & ---                                   \\
		\textbf{\boldmath $^{8}$Li$^{3+}$ } & \textbf{\boldmath 1.653 } & \textbf{\boldmath 0.097 } & \textbf{\boldmath $\beta^-$ } & \textbf{\boldmath $2^+$ $\rightarrow$ $2^+$ } & \textbf{\boldmath 100.0\% } & \textbf{\boldmath 0.840 } & \textbf{\boldmath 6248 } & \textbf{\boldmath $5.8 \cdot 10^{8}$} \\
		$^{16}$N$^{7+}$                     & 1.986                     & 0.127                     & $\beta^-$                     & $2^-$ $\rightarrow$ $6/2^-$                   & 66.2\%                      & 7.130                     & 1942                     & $2.5 \cdot 10^{4}$                    \\
		$^{17}$N$^{7+}$                     & 0.355                     & -0.143                    & $\beta^-$                     & $1/2^-$ $\rightarrow$ $3/2^-$                 & 50.3\%                      & 4.171                     & 1457                     & $1.0 \cdot 10^{5}$                    \\
		$^{20}$F$^{9+}$                     & 2.093                     & 0.154                     & $\beta^-$                     & $2^+$ $\rightarrow$ $2^+$                     & 100.0\%                     & 11.070                    & 2482                     & $9.7 \cdot 10^{6}$                    \\
		$^{12}$B$^{5+}$                     & 1.003                     & 0.196                     & $\beta^-$                     & $1^+$ $\rightarrow$ $0^-$                     & 98.2\%                      & 0.020                     & 6439                     & ---                                   \\
		$^{17}$N$^{5+}$                     & 0.355                     & 0.199                     & $\beta^-$                     & $1/2^-$ $\rightarrow$ $3/2^-$                 & 50.3\%                      & 4.171                     & 1457                     & $1.0 \cdot 10^{5}$                    \\
		$^{17}$Ne$^{10+}$                   & 0.787                     & 0.330                     & $\beta^+$                     & $1/2^-$ $\rightarrow$ $3/2^-$                 & 49.0\%                      & 0.109                     & 3806                     & $4.5 \cdot 10^{3}$                    \\
		$^{15}$O$^{8+}$                     & 0.719                     & 0.338                     & $\beta^+$                     & $1/2^-$ $\rightarrow$ $1/2^-$                 & 99.9\%                      & 122.240                   & 735                      & ---                                   \\
		$^{14}$B$^{3+}$                     & 1.185                     & 0.375                     & $\beta^-$                     & $2^-$ $\rightarrow$ $1^-$                     & 79.0\%                      & 0.012                     & 7024                     & ---                                   \\
		$^{13}$N$^{3+}$                     & 0.322                     & 0.385                     & $\beta^+$                     & $1/2^-$ $\rightarrow$ $1/2^-$                 & 99.8\%                      & 597.900                   & 492                      & $2.4 \cdot 10^{4}$                    \\
		$^{19}$O$^{4+}$                     & 1.532                     & 0.445                     & $\beta^-$                     & $5/2^+$ $\rightarrow$ $3/2^+$                 & 54.4\%                      & 26.880                    & 1442                     & $1.3 \cdot 10^{5}$                    \\
		$^{22}$F$^{5+}$                     & 2.694                     & 0.471                     & $\beta^-$                     & $4^+$ $\rightarrow$ $4^+$                     & 53.9\%                      & 4.230                     & 2449                     & $2.5 \cdot 10^{4}$                    \\
		$^{20}$F$^{7+}$                     & 2.093                     & 0.484                     & $\beta^-$                     & $2^+$ $\rightarrow$ $2^+$                     & 100.0\%                     & 11.070                    & 2482                     & $9.7 \cdot 10^{6}$                    \\
		$^{13}$O$^{4+}$                     & 1.389                     & 0.497                     & $\beta^+$                     & $3/2^-$ $\rightarrow$ $1/2^-$                 & 89.2\%                      & 0.009                     & 8136                     & ---                                   \\
		$^{21}$Na$^{11+}$                   & 2.386                     & 0.507                     & $\beta^+$                     & $3/2^+$ $\rightarrow$ $3/2^+$                 & 94.8\%                      & 22.490                    & 1110                     & $4.0 \cdot 10^{8}$                    \\
		$^{16}$N$^{5+}$                     & 1.986                     & 0.577                     & $\beta^-$                     & $2^-$ $\rightarrow$ $6/2^-$                   & 66.2\%                      & 7.130                     & 1942                     & $2.5 \cdot 10^{4}$                    \\
		$^{25}$Na$^{11+}$                   & 3.683                     & 0.661                     & $\beta^-$                     & $5/2^+$ $\rightarrow$ $5/2^+$                 & 62.5\%                      & 59.100                    & 1714                     & $2.6 \cdot 10^{9}$                    \\
		$^{17}$Ne$^{8+}$                    & 0.787                     & 0.662                     & $\beta^+$                     & $1/2^-$ $\rightarrow$ $3/2^-$                 & 49.0\%                      & 0.109                     & 3806                     & $4.5 \cdot 10^{3}$                    \\
		$^{17}$F$^{9+}$                     & 4.721                     & 0.770                     & $\beta^+$                     & $5/2^+$ $\rightarrow$ $5/2^+$                 & 99.9\%                      & 64.490                    & 739                      & $1.1 \cdot 10^{7}$                    \\
		$^{15}$O$^{6+}$                     & 0.719                     & 0.785                     & $\beta^+$                     & $1/2^-$ $\rightarrow$ $1/2^-$                 & 99.9\%                      & 122.240                   & 735                      & ---                                   \\
		$^{21}$F$^{9+}$                     & 3.919                     & 0.815                     & $\beta^-$                     & $5/2^+$ $\rightarrow$ $5/2^+$                 & 74.1\%                      & 4.158                     & 2452                     & $9.4 \cdot 10^{5}$                    \\
		$^{20}$Na$^{+}$                     & 0.369                     & 0.834                     & $\beta^+$                     & $2^+$ $\rightarrow$ $2^+$                     & 79.3\%                      & 0.448                     & 5392                     & $1.1 \cdot 10^{6}$                    \\
		$^{21}$Na$^{9+}$                    & 2.386                     & 0.842                     & $\beta^+$                     & $3/2^+$ $\rightarrow$ $3/2^+$                 & 94.8\%                      & 22.490                    & 1110                     & $4.0 \cdot 10^{8}$                    \\
		$^{12}$B$^{3+}$                     & 1.003                     & 0.994                     & $\beta^-$                     & $1^+$ $\rightarrow$ $0^-$                     & 98.2\%                      & 0.020                     & 6439                     & ---                                   \\
		$^{17}$N$^{3+}$                     & 0.355                     & 0.999                     & $\beta^-$                     & $1/2^-$ $\rightarrow$ $3/2^-$                 & 50.3\%                      & 4.171                     & 1457                     & $1.0 \cdot 10^{5}$                    \\
		$^{8}$B$^{+}$                       & 1.036                     & 1.062                     & $\beta^+$                     & $2^+$ $\rightarrow$ $2^+$                     & 88.0\%                      & 0.770                     & 6732                     & $6.4 \cdot 10^{4}$                    \\
		$^{9}$C$^{2+}$                      & 1.391                     & 1.079                     & $\beta^+$                     & $3/2^-$ $\rightarrow$ $3/2^-$                 & 54.1\%                      & 0.127                     & 7502                     & $4.0 \cdot 10^{3}$                    \\
		$^{8}$Li$^{+}$                      & 1.653                     & 2.292                     & $\beta^-$                     & $2^+$ $\rightarrow$ $2^+$                     & 100.0\%                     & 0.840                     & 6248                     & $5.8 \cdot 10^{8}$                    \\
		$^{9}$Li$^{3+}$                     & 3.437                     & 2.421                     & $\beta^-$                     & $3/2^-$ $\rightarrow$ $3/2^-$                 & 49.2\%                      & 0.178                     & 6562                     & $3.6 \cdot 10^{7}$                    \\
		\bottomrule
	\end{tabular}
\end{table}

Table~\ref{tab:isotope_summary} contains all ions with $a < 1$,
showing their mode of decay and the most probable transition. The
reported yield for production at ISOLDE per \textmu C of beam on the
target is also given to give a sense of abundance. The typical mode of
operation at this facility is \SIrange{1}{2}{\micro C} every
\SIrange{1}{2.5}{s}. Note that the yield refers only to $+1$ ions.
Additionally, ions with a negative AMM are omitted as for them the
frozen-spin electric field acts in the opposite direction of the
motional electric field, thus reducing the sensitivity. However, ions
with a small negative $a$ or low ionization state are kept as
potentially interesting candidates.

Among the candidate isotopes listed in
Table~\ref{tab:isotope_summary}, the fully stripped $^8$Li nucleus
($J^\pi = 2^+$, $T_{1/2} = \SI{0.84}{s}$) emerges as one of the most
promising for an EDM measurement. Its favorable characteristics
include a small anomalous magnetic moment, a suitable half-life for
storage-based experiments, and high production yield. The decay is a
pure $\beta^-$ transition with a 100\% branching ratio
($E_{\mathrm{max}} = \SI{13.1}{MeV}$) to an excited state of $^8$Be,
which promptly disintegrates into two $\sim \SI{1.5}{MeV}$ alpha
particles. These decay products can be easily stopped if necessary and
may serve as useful beam diagnostics during storage. The beta-decay
asymmetry parameter for $^8$Li is $\alpha = 1/3$.

\subsection{Experimental sensitivity}

The most favorable lithium ion, $^8$Li$^{3+}$, is experimentally
challenging to obtain because a polarized beam of fully stripped
nuclei is currently not available. However, the single charged state
is a common beta-NMR probe with reported polarization levels up to
69\%~\cite{LEVY2002}. Although it has a relatively large anomalous
magnetic moment ($a = 2.29$), $^8$Li$^{+}$ can serve as a precursor to
validate key techniques with single charged ions, albeit with reduced
sensitivity.

Using the frozen-spin trap developed for the muon EDM
search~\cite{Adelmann2025} we will demonstrate the feasibility using
single charged ions. The nominal orbit radius in the experiment is
fixed to \SI{30}{mm} and the maximum achievable electric field is of
the order of \SI{2}{MV/m}~\cite{Hume2024}. The solenoid can produce
fields of up to \SI{5}{T}, however, longitudinal beam injection
requires shielding of the fringe fields. Magnetic fields up to about
\SI{1}{T} can be shielded with steel or iron, while stronger fields
necessitate superconducting channels. To simplify a proof-of-concept
experiment, operation at lower fields is preferable.

\begin{figure}[htbp]
	\centering \includegraphics[angle=0,width=1.0\textwidth]{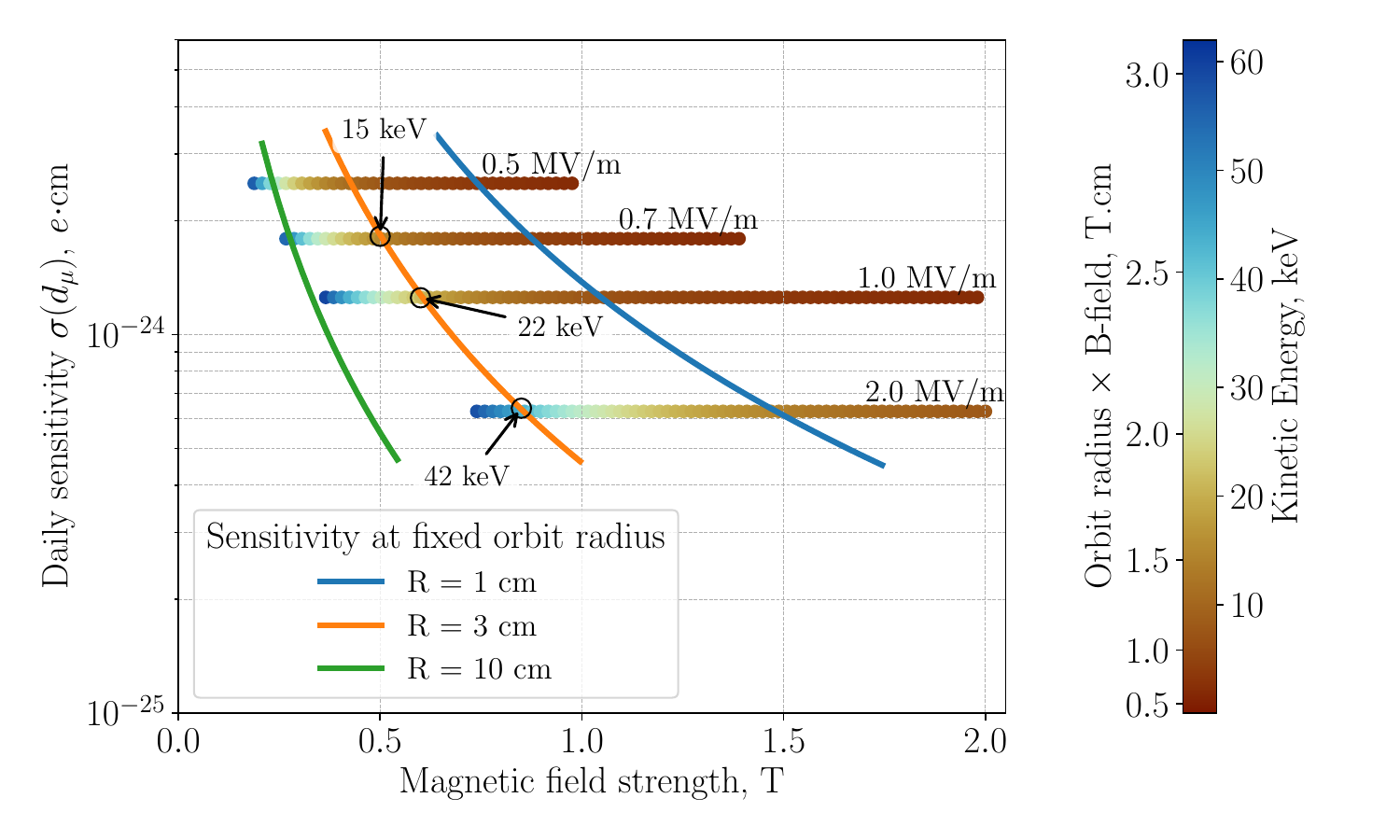}
	\caption{Daily sensitivity to the EDM of singly charged $^8$Li ions
		as a function of the strength of the confining magnetic field. The
		color gradient shows the kinetic energy of the particles for a
		given frozen-spin electric field. The three solid lines show the
		sensitivity at a fixed value of the orbit radius. A storage time
		of \SI{1}{s} is assumed.}
	\label{fig:_figures_sensitivity_li8}
\end{figure}

Figure~\ref{fig:_figures_sensitivity_li8} shows the estimated daily
sensitivity to the EDM of $^8$Li$^+$ as a function of the key
parameters: applied electric and magnetic fields, beam kinetic energy,
and orbit radius inside the ion trap. In the non-relativistic
approximation, applicable for the discussed energy scale where
$\left(\gamma-1\right) < \num{5e-6}$, the relationship between the
magnitude of the frozen-spin field, the orbit radius, and the magnetic
field is %
\begin{equation}
	\lvert E_\mathrm{f} \rvert = \frac{q}{m}a(a+1)B^2 R.
	\label{eq:simple_frozen_spin}
\end{equation}
The required kinetic energy for this condition is %
\begin{equation}
	\label{eq:kinetic_energy}
	K = \frac{\left( q\left(a+1\right) B R \right)^2}{2m}.
\end{equation}
Note that, from Eq.~\eqref{eq:magnetic_anomaly}, it follows that
$q(a+1)$ is approximately constant for a given isotope. Consequently,
the required electric field scales linearly with the magnetic anomaly
$a$, while remaining independent of the ion’s charge state. As a
result, the kinetic energy also remains unaffected by the charge
state.

Equations~\eqref{eq:stat_sens_per_p} and \eqref{eq:number_decayed}
were used to estimate the daily sensitivity in
Fig.~\ref{fig:_figures_sensitivity_li8}. The trapping efficiency,
$\epsilon = 1\%$, is estimated by scaling the muonEDM efficiency due
to the much lower emittance of the ion beamline
(\SI{6}{mm\ mrad}\footnote{ISOLDE Newsletter 2023, CERN, \href{https://isolde.cern/sites/default/files/ISOLDE\%20Newsletter\%202023_0.pdf}{https://isolde.cern/},
	accessed on 3 June 2025.}) compared to the muon beamline
(\SI{200}{mm\ mrad}). Given the significantly smaller phase space and
energy spread of the ion beam compared to the muonEDM experiment at
PSI, a higher injection efficiency is expected, corresponding to
\num{5.6e6} particles per fill every \SI{5}{s}.

The three points in Fig.~\ref{fig:_figures_sensitivity_li8} show
possible configurations of the LionEDM experiment at \SI{3}{cm} orbit
radius, all possible to implement in the muonEDM apparatus. The
$K=\SI{42}{keV}$ point corresponds to \SI{6.4e-25}{\ecm} daily or
\SI{2.4e-25}{\ecm} weekly sensitivity. The latter corresponds to
\SI{20}{\micro rad / s} spin precession.

The velocity of the beam in this configuration is $\beta c \approxeq
	\num{3.4e-3}c$ or \SI{1000}{ mm / \micro s}. After exiting the
injection channel, the beam will spiral towards the storage area with
a longitudinal velocity \SI{10}{mm/\micro s}, corresponding to
\SI{10}{mrad} pitch of the helix.

The particle beam can then be decelerated and stopped in the center by
an electric field gradient of about \SI{10}{V/m} generated by a
Penning-trap-like configuration, shown in Fig.~\ref{fig:psc_setup}.
Once confined, the beam would orbit in the \SI{2}{MV/m} radial
electric field generated by concentric cylindrical electrodes. The
high-voltage required to generate this field strength is \SI{20}{kV},
keeping a spacing of \SI{10}{mm} between the electrodes.

The trajectory of the decay products coming from the stored nuclei
will be deduced from scintillating trackers placed around the outer
electrode. The mean beta-decay energy of \SI{6.2}{MeV} for $^8$Li
results in electron tracks within a \SIrange{0.5}{1}{T}-field that are
similar in spatial distribution as the positrons tracks from muon
decay, for which the muonEDM experiment was optimized. The latter is
conducted at a higher magnetic field of \SI{2.5}{T} and detects
positrons in the range of \SIrange{25}{66}{MeV}. The need for a
tracker outside the outer electrode places stringent constraints on
its thickness and material composition as the decay particles need to
escape and, ideally, make a few turns before stopping in the material.

Electrons emitted in the lower energy region of the beta-decay
spectrum, as well as electrons with a low transverse momentum, will be
detected by end-cap detectors placed up- and downstream of the
electrode system. These end-cap detectors are sufficient to collect
all decay particles for the case that $^8$Li$^{3+}$ becomes available
as then one would run the experiment at a much higher magnetic field
where the electron tracks are confined in the volume between the
electrodes.

Storing the low-energy ion beam for a long period of time requires
ultra-high vacuum. At the level of \SI{e-8}{mbar}, the ions would lose
close to 1\% of their kinetic energy in \SI{1}{s} of storage time.
Reaching this level of vacuum requires differential pumping to
separate the experiment from the beamline, which is typically at
\SI{e-6}{mbar}.

To freeze the spin-motion due to the anomalous magnetic moment, the
electric field must be precisely tuned. Nevertheless, a spread in the
kinetic energy of the particles in the bunch leads to a gradual loss
of polarization and reduction of sensitivity. The anomalous spin
precession frequency %
\begin{equation}
	\label{eq:amm_nonrel}
	\omega_a = -\frac{q}{m}\left(aB - \frac{E}{\beta c}\right)
\end{equation}
depends on the electric and magnetic field experienced by the
individual particle. While the solenoidal magnetic field is constant
throughout the storage volume, the radial electric field strength
generated by the concentric electrodes decreases with $1/R$ following
Gauss's law. Therefore, particles with higher than nominal velocity
experience a lower electric field, contrary to the required higher
electric field that would be needed to satisfy
\eqref{eq:FrozenSpinCondition}. This will lead to a gradual dephasing
of the spin orientation as a function of the energy spread. Taking the
derivative of \eqref{eq:amm_nonrel} with respect to the kinetic
energy, we obtain the spread of $\omega_a$,
\begin{equation}
	\label{eq:spread_omega_z}
	\sigma(\omega_a) = a \frac{qB}{m} \delta(K),
\end{equation}
as a function of the relative spread in $K$, where $\delta(K) =
	\sigma(K)/K$. The spin-coherence time is then approximately $T_s = 2
	\pi / \sigma(\omega_a)$.

For ion beams originating from surface ionization or laser ablation
sources, the relative energy spread is typically $\delta(K) \approxeq
	\num{1e-4}$, i.e.\ several \unit{eV}. This corresponds to a spin
coherence time of only $T_s = \SI{2.7}{ms}$ for $^8$Li$^{+}$ ions
stored at \SI{42}{keV} kinetic energy in a \SI{0.85}{T} magnetic
field. Such a coherence time is far shorter than the possible ion
storage time. Moreover, a finite beam divergence contributes an
additional spread in orbital kinetic energy. A transverse divergence
$\theta$ changes the orbital velocity fractionally by $\theta^2 / 2$,
and hence the orbital kinetic energy by $\theta^2$. To achieve a
\SI{1}{s} spin coherence time, one requires $\theta^2 \leq
	\num{3e-7}$, i.e.\ $\theta \leq \SI{0.5}{mrad}$.

From these constraints we can infer the necessary input beam quality
in terms of emittance. The longitudinal emittance $\xi_\mathrm{long}
	\approx \pi \, \Delta E \, \Delta t$ is given by the product of the
bunch temporal width with its energy spread. For a successful
measurement one requires $\xi_\mathrm{long} \approxeq
	\SI{0.013}{eV.\micro s}$, corresponding to $\Delta E = \SI{12}{meV}$
and $\Delta t = \SI{1}{\micro s}$. The transverse emittance should be
limited to $\xi_\mathrm{trans} < \SI{0.5}{mm.mrad}$ in order to
suppress beam divergence, extend spin coherence, and maintain a spot
size below \SI{1}{mm} for efficient injection into the trap channels.

Direct injection from the ion source is therefore not feasible due to
its intrinsically large emittance. Prior to delivery to the
frozen-spin ion trap, the beam must be cooled and bunched. For that
purpose, room-temperature radio-frequency quadrupole cooler-bunchers
(RFQcb) are commonly used in radioactive ion beam facilities, using
buffer-gas cooling. One such RFQcb is already operational at the
high-resolution spectrometer beamline at
CERN-ISOLDE~\cite{Catherall2017}, however, its extraction emittance is
too large for a light-ion EDM experiment. A cryogenic implementation,
by contrast, can provide the required phase space compression.
Recently, Lechner et al.\ demonstrated in simulations
$\xi_\mathrm{long} = \SI{0.02}{eV.\micro s}$ and $\xi_\mathrm{trans} =
	\SI{0.02}{\pi.mm.mrad}$ for $^{20}$Mg$^+$ ions using a cryogenic Paul
trap with He buffer gas at \SI{40}{K}, which is being developed in the
context of the Multi Ion Reflection Apparatus for Collinear Laser
Spectroscopy (MIRACLS) also at CERN-ISOLDE~\cite{Lechner2024}. For
$^8$Li$^+$, an even smaller transverse emittance can be expected,
scaling approximately with the square root of the mass ratio between
lithium and magnesium.

Another possibility that will be explored to achieve long term spin
coherence is to apply dedicated techniques like those used at COSY
where a \SI{1000}{s} proton beam polarization lifetime was
reported~\cite{Guidoboni2016}, or spin-echo pulse schemes to cancel
relative dephasing similar as in Ref.~\cite{Afach2015PRL}.

Overall, the $^8\mathrm{Li}^{+}$ ion is an ideal candidate for this
pioneering study due to its well-established high-polarization
technique and simple nuclear structure, which facilitates
interpretation of results. Even at low magnetic fields, a single day
of measurement could achieve \SI{e-24}{\ecm} sensitivity, which
approaches the best indirect proton EDM limits. The important
parameters of the experiment are summarized in
Table~\ref{tab:li8_parameters}.

\begin{table}[htbp]
	\centering
	\caption{Key parameters for a $^8$Li$^{+}$ proof-of-concept EDM
		experiment.}
	\label{tab:li8_parameters}
	\begin{tabular}{l l}
		\hline
		Anomalous magnetic moment $a$             & 2.29                                           \\
		Mean $\beta$-decay energy                 & \SI{6.2}{MeV}                                  \\
		Nominal orbit radius $R$                  & \SI{30}{mm}                                    \\
		Kinetic energy $K$                        & \SI{42}{keV}                                   \\
		Beam velocity $\beta c$                   & $3.4\times 10^{-3} c$ (\SI{1000}{mm/\micro s}) \\
		Frozen-spin electric field $E_\mathrm{f}$ & \SI{2}{MV/m}                                   \\
		Confining solenoidal $B$-field            & \SI{850}{mT}                                   \\
		Input beam emittance                      & $\xi_\mathrm{long} \leq$
		\SI{0.01}{eV.\micro s}; $\xi_\mathrm{trans} \leq \SI{0.5}{mm.mrad} $                       \\
		Particles per fill                        & $5.6\times 10^6$ every \SI{5}{s}               \\
		Storage time                              & \SI{1}{s}                                      \\
		Vacuum requirement                        & $<\SI{e-8}{mbar}$                              \\
		Weekly EDM sensitivity                    & \SI{2.4e-25}{\ecm}                             \\
		Corresponding spin precession             & \SI{20}{\micro rad/s}                          \\
		\hline
	\end{tabular}
\end{table}

Significant improvements in EDM sensitivity can be achieved using
fully stripped nuclei. The $^8$Li$^{3+}$ state offers an eightfold
enhancement due to its 24 times lower AMM compared to $^8$Li$^{+}$,
following the $a/(a+1)$ scaling in Eq.~\eqref{eq:stat_sens_per_p}. A
further threefold improvement arises from reduced electron shielding,
described by the $Z/z$ factor in Eq.~\eqref{eq:stat_sens_per_p}. If
the experiment is limited by the spin-coherence time, sensitivity
improves linearly with $aq$, yielding an additional eightfold
enhancement.

When transitioning from single charged to fully stripped ions
statistical losses might occur due to the electron stripping process.
Nonetheless, the resulting sensitivity improvement ranges from two to
three orders of magnitude, making fully stripped $^8$Li a highly
promising probe for future EDM searches. Moreover, the systematic
effects arising from the coupling of the AMM to the experiment’s
electromagnetic fields would be reduced by a factor of 24, further
enhancing sensitivity limits.


\subsection{Systematic effects and countermeasures}

Reaching high sensitivity to light ion EDMs requires a stringent
control of spurious effects that mimic a true EDM signal. The dominant
systematic effects in a frozen-spin trap were originally laid out in
\cite{Farley2004PRL} and later elaborated specifically for the muonEDM
experiment in \cite{Cavoto2024}. The most critical systematic effect
is the presence of a net radial magnetic field component, $\langle
	B^*_r \rangle \neq 0$, in the rest frame of the particle. It can be
generated by an axial E-field component $E_z$ or a radial B-field
component, $B_r$, in the laboratory frame. Both field components exert
Lorentz force on the ions, and so, the mean axial force in the
laboratory frame
\begin{equation}
	\label{eq:trapped_requirement}
	q\langle E_z + c \beta B_r \rangle = 0,
\end{equation}
for the orbit to be stable axially.

In a purely magnetic or purely electric confinement $\langle B^*_r
	\rangle = 0$ for trapped particles, as otherwise there will be a net
force which accelerates them out of the storage region, and so there
would be no systematic effect~\cite{Farley2004PRL}. However, one
cannot completely eliminate stray fields and in the general case the
ions will be trapped in a combination of electric and magnetic fields.
Here we will examine the motion of ions in the frozen-spin trap to
estimate the necessary level of control of the electromagnetic fields
in the experiment to achieve the target sensitivity.

The radial magnetic field around the center of the solenoid can be
approximated to first order by a magnetic field gradient $B_r(z, t) =
	z(t) \partial_z B_r$, where $\partial_z \equiv \partial / \partial z$.
Similarly, the electrostatic axial trapping potential can be
approximated by $E_z(z, t) = (z(t) - \Delta_z) \partial_z E_z$, where
$\Delta_z$ is the separation between the zero crossings of the E-field
and B-field gradients, and accounts for the possibility that the
confining electrodes are not ideally symmetrically placed around the
center of the solenoid.

The axial motion of an ion is then
\begin{equation}
	\label{eq:axial_motion}
	m \ddot z(t) = q \Big(\partial_z E_z \big(z(t) - \Delta_z\big) + c
	\beta \partial_z B_r z(t)\Big),
\end{equation}
where $c \beta = \sqrt{2K/m}$ is the velocity of the particle along
its orbit. The solution of Eq.~\eqref{eq:axial_motion}, assuming $z(0)
	= z_0$, is
\begin{equation}
	\label{eq:axial_motion_sol}
	z(t) = \frac{q}{m} \frac{\partial_z E_z (z_0 - \Delta_z) + c \beta
		\partial_z B_r z_0}{\varpi^2} \cos(\varpi t) +
	\frac{q}{m}\frac{\partial_z E_z \Delta_z}{\varpi^2},
\end{equation}
a harmonic oscillator with angular frequency $\varpi= \sqrt{-q/m
		(\partial_z E_z + c \beta \partial_z B_r)}$. Note that negative
field gradients yield a real frequency, and therefore confinement. The
constant term
\begin{equation}
	\label{eq:equilibrium_z}
	z_\mathrm{eq} = \frac{q}{m}\frac{\partial_z E_z \Delta_z}{\varpi^2}
\end{equation}
is the $z$-position of the equilibrium of this harmonic oscillator.

In the absence of the constant term in
Eq.~\eqref{eq:axial_motion_sol}, the particle will oscillate around
zero, and thus, if there is no spatial separation $\Delta_z$ between
the axially confining E-field and the center of the solenoid, no
systematic effect will be generated. If the fields are not overlapping
ideally, a net spin rotation
\begin{equation}
	\label{eq:false_edm_bfield}
	\omega_\mathrm{f} = \frac{q}{m} \left(a \partial_z B_r z_\mathrm{eq}
	- \frac{\partial_z E_z \left(z_\mathrm{eq} - \Delta_z\right)}{c
		\beta}\right)
\end{equation}
will be observed.

Substituting \eqref{eq:equilibrium_z} into
\eqref{eq:false_edm_bfield}, the resulting EDM-like precession
frequency can be expressed as
\begin{equation}
	\label{eq:false_edm_bfield_final}
	\omega_\mathrm{f} \simeq \frac{q}{m}\, a\, \partial_z B_r\,
	\Delta_z,
\end{equation}
as a function of the separation $\Delta_z$ between the electric and
magnetic field zero crossings, assuming $c\beta\, \partial_z B_r \ll
	\partial_z E_z$. The latter is already required to suppress systematic
effects by ensuring sufficiently small magnetic-field gradients.

A magnetic-field uniformity with gradients below $\SI{30}{nT/mm}$ was
demonstrated in Fermilab’s muon $g-2$ storage ring
experiment~\cite{Albahri2021, FNAL2025PRL}. Achieving a similar level
of shimming for the main solenoid and constraining $\Delta_z \leq
	\SI{0.1}{\micro m}$, the false spin precession of $^8$Li$^+$, would be
limited to less than the required $\SI{20}{\micro rad/s}$. However,
direct measurement of $\Delta_z$ to that precision is experimentally
challenging. To control this systematic contribution
$\omega_\mathrm{f}$ and isolate the true EDM signal, we follow a
strategy analogous to the crossing-point analysis used in the nEDM
experiment that placed the most stringent upper limit on the neutron
EDM with a sensitivity of $\SI{1.8e-26}{\ecm}$ (90\%
C.\@L.\@)~\cite{Abel2020}.

In the presence of a permanent EDM, the measured precession frequency
takes the form
\begin{equation}
	\label{eq:cross_technique_mesured}
	\omega_\mathrm{m} = \omega_\mathrm{f} + \omega_e = \frac{q}{m}a
	\partial_z B_r \Delta_z + \omega_e,
\end{equation}
which is a linear relation of the form $y = kx + b$ with $x \equiv
	\Delta_z$. Measuring $\omega_\mathrm{m}$ as a function of $\Delta_z$
therefore yields a straight line with slope directly related to the
axial gradient of the radial magnetic field $\partial_z B_r$. The
separation $\Delta_z$ can be varied by adjusting the voltages on the
axial confinement electrodes surrounding the grounded ring (see
Fig.~\ref{fig:psc_setup}).

The \CP- and $T$-violating character of the EDM enables a
time-reversal-symmetric measurement. This is implemented by reversing
the currents in all coils, thereby inverting the magnetic-field
direction, and by switching the beam propagation from CW to CCW. A
second measurement is then performed under the inverted magnetic-field
gradient, again scanning over $\Delta_z$. This yields another straight
line with a modified slope but the same intercept, $\omega_e$. The two
lines intersect at $\Delta_z = 0$, where the systematic effect
cancels, leaving the observed spin-precession frequency equal to the
EDM-induced value $\omega_e$.

Another potential systematic effect in the experiment could arise from
oscillations of the electromagnetic field in the ion’s rest frame.
Small fluctuations around the nominal field along perpendicular axes
would induce spin rotations; since such rotations do not commute,
their interplay can generate a net rotation that mimics an EDM signal.
For instance, ions traveling slightly off the nominal orbit experience
an oscillating electric field around the frozen-spin value. This can
couple to an oscillating magnetic-field component along the ion
trajectory, leading to a gradual phase accumulation. This effect is an
example of Berry’s phase~\cite{Berry1984}.

Systematic effects of this type were studied in detail for the muon
EDM experiment~\cite{Cavoto2024}. While they were found negligible in
that context, they may become relevant for light-ion experiments due
to higher sensitivity and longer observation times. Nonetheless,
several mitigating factors exist. Phase accumulation is significant
only for resonant oscillations along two perpendicular axes, which
greatly restricts possible error sources. Moreover, the effect cancels
to first order when using the CW/CCW injection scheme. Finally, the
spin’s oscillation frequencies and amplitudes around different axes
can be reconstructed from the measured decay products of the ions.
This information enables calculation and correction of the EDM-like
phase shift.


\subsection{Sensitivity to fundamental \textit{CP}-odd parameters}

Results from EDM searches using light ions in a compact storage trap
can be related to the fundamental \CP-odd parameters. 
In this subsection, we address the physics potential of measuring the
EDMs of ${\rm ^8Li}$ and ${\rm ^9Li}$ ions employing an intermediate
model of \CP violation parametrized in terms of the nucleon EDMs,
$\{d_p,d_n\}$. While the neutron EDM limits are well-known
\cite{Abel2020}, the indirect limits on the proton EDM can be
extracted from atomic EDM measurements
\cite{Dmitriev:2003sc,Flambaum:2019ejc}. Here we use the results of
Refs. \cite{Dmitriev:2003sc} that relate the experimental results on
the EDM of ${\rm ^{199}Hg}$ atom \cite{Graner:2016ses} with $d_p$ and
$d_n$, %
\begin{eqnarray}
	\left| d_n + \frac{0.2\pm 0.02}{1.9} d_p\right| &<& 1.6 \times 10^{-26} \,{e\rm cm},\nonumber\\|d_n|&<& 1.8 \times 10^{-26} \,{e\rm cm}
\end{eqnarray}
Due to the absence of the valence protons on the outer shell of ${\rm
			^{199}Hg}$, the sensitivity to $d_p$ is markedly reduced. The
combination of the neutron and mercury EDM experiments significantly
constrains $d_n$ but produces a limit on $d_p$ that is considerably
weaker\footnote{Moreover, the presence of $T,P$-odd nuclear forces
	could lead to a further relaxation of a strict limit on $d_p$.}. On
a two-dimensional $\{d_p,d_n\}$ plane, neutron and mercury EDM
constraints can be represented by the two almost parallel bands, see
Fig.~\ref{fig:li8_pedm_pdf}.


In contrast to ${\rm ^{199}Hg}$, the nucleus of ${\rm ^{8}Li}$ ion
contains five neutrons and three protons, and both $d_n$ and $d_p$
will contribute to the nuclear EDM. Employing a naive shell model,
$^8$Li is treated as a $^4$He core plus four valence nucleons—three
neutrons and one proton—occupying $p$-wave orbitals. The single
particle configuration should be $(p_{3/2})^3_n (p_{3/2})^1_p$. If all
four nucleons were neutrons in the $(p_{3/2})$ shell,
antisymmetrization would require them to fill all $m_j$ sublevels,
resulting in a $J = 0$ state. Thus, the $(p_{3/2})^3_n$ configuration
corresponds effectively to a single neutron hole in the $p_{3/2}$
shell, giving total angular momentum $J = 3/2$. For the EDM problem,
this neutron hole can be thought of as an actual neutron in the
$p_{3/2}$ state, and its magnetic moment and EDM will be same as for a
real neutron. The same conclusion could be made for the proton,
effectively giving a $(p_{3/2})^1_n(p_{3/2})^1_p$ state.

Hence, to a good approximation, the unpaired nucleons in $^8$Li are
one neutron and one proton, both in $p_{3/2}$ states. When combining
two angular momenta $j = 3/2$, the total angular momentum can range
from $J = 0$ to $J = 3$. The observed $J^\pi = 2^+$ ground state
arises from more complex configuration mixing, but for an EDM
estimate, we can construct the $|J = 2, M = 2\rangle$ state directly
from coupled $|3/2, m\rangle_n |3/2, m'\rangle_p$ states. This
approach is limited and in a more realistic model, there would be
admixtures of other states like $p_{1/2}$. Nevertheless, we can
evaluate the magnetic moment of $^{8}$Li in this simplified model to
estimate its validity.

Using vacuum values for the spin and orbital $g$-factors of the proton
and neutron, this simple model yields a magnetic moment of $\mu
	\approx 1.25~\mu_N$ for $^8$Li, which compares favorably to the
shell-model prediction of 1.366 by Cohen and Kurath as cited in
Ref.~\cite{vanHees1988} and the experimental value of $1.65\mu_N$.
While this $\approx25\%$ deviation reflects the crudeness of the
model, it is sufficient to evaluate the sensitivity to the neutron and
proton EDMs in a lithium experiment. 

Within the same simplistic model we can calculate the EDM of $^{8}$Li.
To do that we assume that the ground state of lithium nucleus is fully
polarized, and is in the $\lvert J = 2, M = 2\rangle$ state. This
gives
\begin{equation}
	d_{^8\mathrm{Li}} = \langle 2, 2 | d_p \sigma_{pz} + d_n \sigma_{nz}
	| 2, 2 \rangle = \frac{2}{3}(d_n + d_p) \,,
	\label{eq:li8_theory_edm}
\end{equation}
where $\sigma_{pz}$ and $\sigma_{nz}$ are the spin operators of the
valence proton and neutron projected along the quantization axis.

Relation~\eqref{eq:li8_theory_edm} demonstrates that $^8$Li has
significant sensitivity to both $d_n$ and $d_p$. For comparison, the
sensitivity to $d_p$ is substantially higher than that of $^{199}$Hg.
As illustrated in Fig.~\ref{fig:li8_pedm_pdf}, an experimental
sensitivity at the level of $\mathcal{O}(10^{-25})~e\,\mathrm{cm}$
would surpass existing constraints on the proton EDM.

Moreover, these results can be made considerably more precise using
the {\em ab-initio} techniques of the nuclear theory applicable to low
$A$ nuclei. Such methods have already been successfully applied to
calculate the EDMs of $^6$Li, $^9$Be, and
$^{13}$C~\cite{Yamanaka2017}, as well as $^7$Li and
$^{11}$B~\cite{Yamanaka2019}. In the future, one can also include the
influence of the \CP-odd nuclear forces on EDMs of lithium isotopes,
and it is expected to be a more feasible task than a corresponding
calculation for the mercury EDM, where it remains to be rather
difficult \cite{Engel2025}.

Light nuclei can also be treated from first principles using effective
field theory, similar to the deuteron~\cite{Lebedev2004}. As shown in
Ref.~\cite{Dekens2014}, EDM measurements of the deuteron and helion in
storage rings can discriminate between flavor-diagonal sources of
\CP\ violation, including the QCD $\theta$-term, the minimal
left-right symmetric model, and related scenarios. Furthermore, $P$-
and $T$-violating nuclear forces can be systematically derived within
chiral effective field theory, which provides a link between
quark-level interactions and nuclear observables~\cite{deVries2020}.
These approaches could also be extended to $^8$Li, offering enhanced
sensitivity to physics beyond the SM.

\section{Conclusions}

The measurement of EDMs in light ions confined in compact traps
represents an exciting opportunity in the ongoing search for new
sources of \CP violation. Current state-of-the-art EDM experiments
predominantly target heavy neutral atoms such as $^{199}$Hg,
$^{225}$Ra, or molecular systems like $^{205}$TlF, or rely on
large-scale infrastructure such as storage rings operating at the
magic momentum for protons and deuterons. These approaches, while
powerful, are inherently limited by theoretical complexity or scale.

In contrast, light ion systems offer a complementary and potentially
more accessible path. Among these, we show that the fully stripped
$^8$Li ion is a uniquely promising candidate. Its low anomalous
magnetic moment, favorable nuclear structure, and accessibility to
first-principles theory make it particularly well suited for EDM
searches in a frozen-spin trap. We estimate a sensitivity of
\SI{1e-26}{\ecm} within a single week of data collection, assuming an
ion yield of \num{5.6e6} ions every \SI{5}{s}.

While the production of fully stripped, polarized $^8$Li beams remains
a technical challenge, the singly charged $^8$Li ion---widely used in
$\beta$-NMR applications---offers a practical and immediate path
toward a proof-of-principle experiment. Remarkably, such a precursor
study could surpass current proton EDM limits derived from $^{199}$Hg
with only a few days of measurement time.

Beyond the immediate prospects of $^8$Li, the successful
implementation of a frozen-spin ion trap would enable precision
measurements across a broad range of light ions, each probing distinct
facets of \CP symmetry violation, and would give access to a new path
toward uncovering physics beyond the Standard Model.

\begin{acknowledgments}
	The authors gratefully acknowledge Y. Stadnik (University of Sydney)
	for the insightful discussions on the theoretical aspects of this
	work. The authors also thank G. Neyens and N. Severijns (KU Leuven),
	as well as M. Kowalska (CERN, ISOLDE), for valuable input regarding
	beamline configurations and laser polarization techniques. Finally,
	we acknowledge the muonEDM collaboration for their continued support
	and collaboration.

	This work was supported by the Swiss National Science Fund under
	grant \textnumero~204118; the European Union's Horizon 2020 research
	and innovation programme under the Marie Sk\l{}od\-owska-Curie grant
	agreement \textnumero~884104 (PSI--FELL\-OW--III-3i); the Swiss
	State Secretariat for Education, Research and Innovation (SERI)
	under grant \textnumero~MB22.00040; and ETH Zürich under grant
	\textnumero~ETH-48 18-1.
\end{acknowledgments}

\bibliography{muonEDM}
\end{document}